# Itinerant metamagnetism in manganites caused by the field-induced electronic nematic order[*]

Tian Gao[1], Shixun Cao[1,2], Shujuan Yuan[1,2], Chao Jing[1], Baojuan Kang[1], Jincang Zhang[1]

[1]Department of Physics, Shanghai University, Shanghai 200444, China

[2]Institute of Low-dimensional Carbons and Device Physics, Shanghai University, Shanghai 200444, China

**Abstract**

Itinerant metamagnetism transition is observed and studied in perovskite $La_{1-x}Ca_xMn_{0.90}Cu_{0.10}O_3$ system for $x = 0.30$. At a constant low temperature, $10\ K \leq T < 150\ K$, there is a continuous second-order metamagnetism jump from a low magnetic state to a high one with the magnetic field $H$ increasing. However, at an exceeding low temperature, $T = 2.5\ K$, the metamagnetism jump at $H = 3.5\ T$ becomes to be a robust first-order transition, and another metamagnetism transition occurs at a higher field $H = 7.0\ T$. Since there is no charge ordering sign in the present system, it can not be understood by using the phase separation model or the prior martensite/austenitic phase transition scenario. A theoretical electronic nematic order phase formation is evidenced to answer for the two consecutive metamagnetic transitions, which separate the nematic phase from the low-field ($H < 3.5\ T$) and high-field ($H > 7.0\ T$) isotropic phases.

[*] This work is supported by the National Natural Science Foundation of China (NSFC, No.10674092, 10774097), the Science and Technology Innovation Fund of the Shanghai Education Committee (No.09ZZ95), and the Science & Technology Committee of Shanghai Municipality (No.08dj1400202).

# 1. Introduction

Metamagnetic transition which refers to a jump in magnetization from a low magnetization state to a high magnetization state as an external magnetic field increase, has been observed and studied a lot in strongly correlated electron materials [1-4]. Several models have been employed to explain the attractive phenomenon. In doped NiMn alloys, the magnetization jump usually occurs accompanied with a structure transition which is driven by magnetic field from a martensitic phase to an austenitic phase [5-7]. At the same time, many studies [8-12] indicate that the dynamic phase separation, i.e., the competitive coexistence of ferromagnetic (FM) phase and charge ordering (CO) antiferromagnetic (AFM) phase, plays the crucial role in the process of metamagnetic transition in perovskite manganites at low temperatures. A model system for this study is $La_{5/8-x}Pr_xCa_{3/8}MnO_3$, which is known for its large-scale phase separation and has sub-micrometer-scaled FM and CO-AFM domains proved by transmission electron microscopy [13]. The coexistence of FM and CO reflects the competition between the intrinsic properties of its two starting materials, i.e., FM $La_{5/8}Ca_{3/8}MnO_3$ ($T_C \sim 275$ K) and CO-AFM $Pr_{5/8}Ca_{3/8}MnO_3$ ($T_{CO} \sim 220$ K). The metal-insulator (M-I) transition and the metamagnetic jump steps occur via dynamic percolation of FM phases [14-16], which gives rise to the colossal magnetoresistance.

However, in many recent studies [17], the metamagnetic jump steps are also observed in charge disordering systems, which give more obstacles to the comprehension on phase separation mechanism. As early as 2005, Kee and his coworkers [18-24] developed a theory of itinerant metamagnetism induced by electronic nematic order in manganites, where the spin-dependent Fermi surface instability gives rise to the formation of an electronic nematic order phase upon increasing the applied Zeeman magnetic field. This leads to two consecutive metamagnetic transitions that separate the electronic nematic ordering phase from the low-field and high-field "isotropic" phases.

It is known that $La_{1-x}Ca_xMnO_3$ shows CO insulating phase when $x$ = 0.5-0.9, and a large-scaled strong FM metal phase with double-exchange [25] when $x$ = 0.2-0.4. At $x$ = 0.3, the sample reveals the largest magnetoresistance (MR) effect with the optimum ratio of $Mn^{3+}/Mn^{4+}$. In this work, we observed and studied the metamagnetic transition in perovskite manganites $La_{1-x}Ca_xMn_{0.90}Cu_{0.10}O_3$ system which does not reveal any signal of CO. All the results indicate that the metamagnetic transition occur due to the formation of a electronic nematic ordering phase where only the spin-up (or spin-down) Fermi surface spontaneously breaks the discrete rotational symmetry. And we also give the related experimental evidences and theoretical interpretation other than the phase separation scenario.

## 2. Experimental

The compounds $La_{1-x}Ca_xMn_{0.90}Cu_{0.10}O_3$ (LCMCO, $x$ = 0.05-0.30) were prepared by the standard solid state reaction method. Stoichiometric amounts of $La_2O_3$, $CaCO_3$, $MnO_2$ and CuO powders were mixed, ground and heated at 1100 °C overnight. Then, the powders were pressed into pellets and sintered at 1200°C for 24 h in air, with intermediate grinding. After that the pellets were reground, repressed and sintered at 1300°C for another 24 h. Finally, the samples were slowly cooled down to room temperature with furnace.

Figure 1 shows the quality of the samples characterized by X-ray powder diffraction (XRD, Rigaku 18kW D/max-2550 diffractometer with Cu-K$\alpha$ radiation) at room temperature. All samples show clear single phase of orthorhombically distorted perovskite structure, and no detectable secondary phase is observed. All peaks of the XRD patterns can be indexed with an orthorhombic lattice with the space group *Pbnm*. Temperature dependence of ac susceptibility was measured in the temperature range of 2-300 K using a physical property measurement system (PPMS-9, Quantum Design Inc.). Dc magnetic measurements were performed on a vibrating sample magnetometer (VSM option on PPMS-9). Magnetic hysteresis loops were recorded after the sample was cooled in zero field from room temperature to the required temperatures.

### 3. Results and discussion

Figure 2 shows the temperature dependence of the magnetization under a magnetic field of $H$ = 50 Oe for LCMCO system, following different experimental procedures: zero-field-cooled (ZFC), field-cooled-cooling (FCC), and field-cooled-warming (FCW). In the doping level of $0.05 \leq x \leq 0.20$, the samples display a PM-FM phase transition at the FM ordering temperature $T_C$ which becomes lower with $x$ increases; the FCC curve is almost concurrent with the FCW one entirely at the whole temperature range recorded, while the ZFC curve branches off below $T_C$. For the sample with $x$ = 0.30, a large thermal hysteresis below $T_C$ is observed. That is, the FCC curve diverges from the FCW one below the first ordering transition temperature $T_C \sim 119$ K, and represents another FM ordering transition at a lower temperature which is called $T_C' \sim 57$ K. The warmed ZFC and FCW curves have the same FM ordering temperature $T_C$. Hence, two FM ordering phase transitions are observed in the present LCMCO with $x = 0.30$ sample.

It is clear that the ferromagnetism at low temperatures increases with the increase of Ca content from $x = 0.05$ (the saturation magnetization $M_S \sim 2.5$ emu/g) up to $x = 0.20$ ($M_S \sim 4.0$ emu/g). Therefore, the more $Ca^{2+}$ substitutions induce the more content of $Mn^{4+}$ ions and enhance the double-exchange interactions. However, this effect is less considerable for transport process in LCMCO with $x=0.30$ sample, because the double-exchange along $Mn^{3+}$—$O^{2-}$—$Mn^{4+}$ chain is weakened gradually by the doping of both A- and B-sites, and the ratio of $Mn^{3+}/Mn^{4+}$ is departing from the optimal 7:3. The B-site doped Cu ions segment the long-range FM ordering in $La_{0.7}Ca_{0.3}MnO_3$, which results in the saturation magnetization of LCMCO ($x = 0.30$) sample is only about 2.5 emu/g. The large thermal hysteresis between cooling and warming measurement also appears in AC susceptibility curves (shown in figure 3). $T_C$ and $T_C'$ in both AC and DC curves are totally consistent with each other.

Figure 4 exhibits some typical ZFC isothermal magnetization $M(H)$ curves for the LCMCO ($x = 0.30$) compound, in a cycle sweeping mode ($0 \rightarrow 80$ kOe $\rightarrow 0 \rightarrow -80$ kOe $\rightarrow 0 \rightarrow 80$ kOe). Below $T = 70$ K, a FM state at null field is observed, and a metamagnetic transition occurs at $H < 40$ kOe. The second ascending-field curve fully superposes the descending-field one, indicating that LCMCO ($x = 0.30$) remains in the ferromagnetic state upon removal of the magnetic field. Above $T = 120$ K, the sample shows a PM ground state at zero field and a PM-FM transition is achieved with magnetic field increasing. On the contrary, at $T = 120$ K, the second ascending-field magnetization follows neither the initial magnetization path nor the descending-field one, indicating a reversible PM-FM transition. At $T = 150$ K, the transition is partially reversible, and the second ascending-field magnetization follows the initial magnetization path totally.

As mentioned in the introduction, similar magnetic thermal hysteresis and metamagnetism transitions have been observed and studied in several other phase separated manganite systems [8-12]. They all attribute the robust metamagnetism transition to the high magnetic field energy, which can drive the low CO-AFM state to a high FM state. However, in figure 2 and figure 3, we can not find any CO or AFM transition in the whole temperature range recorded. Therefore, there is no CO-AFM phase in the present system. The metamagnetism behavior can not be understood with the phase separation mechanism. There must be some other deep-seated physical reasons which results in the metamagnetism transition in the system.

According to the electronic nematic order theory [18-24, 26-28], the itinerant metamagnetism can be induced by the formation of electronic nematic order phase which is described by order parameters,

$$\Delta_\alpha = F_2 \langle \hat{Q}^\alpha_{11}(0) \rangle \text{ and } \Delta'_\alpha = F_2 \langle \hat{Q}^\alpha_{12}(0) \rangle \tag{1}$$

where the interaction $F_2$ can be any arbitrary short-ranged interaction, the quadrupolar order parameter is defined using the momentum operators of electrons, i.e., $\hat{Q}_{ij} = \hat{p}_i \hat{p}_j - \frac{1}{2} \hat{p}^2 \delta_{ij}$, $\alpha$ stands for the spin degree of freedom. When a nonzero solution for $\Delta_\alpha$ exists, the spin-$\alpha$ Fermi surface will spontaneously break the discrete rotational symmetry of the lattice and an electronic nematic order phase is formed, inducing metamagnetic transitions.

At a finite temperature close to zero, the free energy $F$ is investigated as a function of the applied magnetic field, $H$, for a given chemical potential $\mu$. A closed Fermi surface is started with. As the external magnetic field, $H$, increases, $\mu_\uparrow$ ($\mu_\downarrow$) increases (decreases) and the spin-up (spin-down) Fermi surface volume increases (decreases). For a small magnetic field, there is no spontaneous breaking of the lattice symmetry in the Fermi surfaces. When $H$ reaches the value that makes the spin-up Fermi surface gets close to the first Brillouin zone boundary, the non-analytic behavior of free energy density $F_\uparrow$ near the van Hove filling leads to the first order transition to the nematic phase or the abrupt deformation of the spin-up Fermi surface. At the same time, the nematic order parameter for the spin-up electrons jumps to a finite value, $\Delta_\uparrow \neq 0$. This represents a first order transition from a closed to an open spin-up Fermi surface. Here, the spin-down Fermi surface is away from the van Hove filling so that the spin-down Fermi surface only changes gradually. As $H$ further increases, another first order transition occurs such that $\Delta_\uparrow$ abruptly jumps down to zero and the lattice symmetry is restored. That is, a nematic phase exists in $H_1 < H < H_2$ and is bounded by two first order transitions from (to) the low (high) field "isotropic" phases.

In order to reach a further insight into the metamagnetism transition behavior, we carried out the magnetic hysteresis loop measurements at an exceeding low temperature, $T = 2.5$ K, as shown in figure 5. Two consecutive metamagnetic transitions which is in good consistent with the theory of electronic nematic order occur at $H_1 = 3.5$ T and $H_2 = 7.0$ T, respectively. At $H_1 = 3.5$ T, the spin-up (or spin-down) Fermi surface gets close to the van Hove filling and breaks the rotational symmetry of the lattice, an nematic phase is formed. Such a transition on the lattice is generically first order and naturally gives rise to metamagnetism with the jump in the magnetization. The first order transition turns into second order at higher temperatures (see FIG. 4) so that the nematic phase is bounded above by second order transition. As the magnetic field further increases up to $H_2 = 7.0$ T, another first order transition happens. In the high field phase, the lattice symmetry is restored and the spin-up Fermi surface becomes closed again. It should be noticed that, in this case, the occurrence of the nematic order is the source of the metamagnetism, which may be independent with the phase separation model.

## 4. Conclusions

Two consecutive metamagnetic transitions are observed in the optimum doped $La_{0.70}Ca_{0.30}Mn_{0.90}Cu_{0.10}O_3$ manganite at exceeding low temperature $T = 2.5$ K. There is no signal of CO phase in the whole temperature range, so that the metamagnetic jumps can not be understood by the coexistence of CO-AFM phase and FM phase, i.e., phase separation scenario. The field-induced transitions can be well explained by the electric nematic order theory, in which the spin-up (or spin-down) Fermi surface spontaneously breaks the rotational symmetry of the lattice and a nematic phase is formed between two isotropic phase. Such a transition on the lattice is generically first order and naturally gives rise to metamagnetism with the jump in the magnetization across the "isotropic" nematic transition. The first order transition turns into second order one at higher temperatures so that the nematic phase is bounded above by second order transition.

Figure Captions

FIG. 1 X-ray-diffraction patterns of $La_{1-x}Ca_xMn_{0.90}Cu_{0.10}O_3$ ($x$ = 0.05-0.30).

FIG. 2 Temperature dependence of dc magnetization for LCMCO system. Curves are measured under zero-field-cooled (ZFC), field-cooled-cooling (FCC), and field-cooled-warming (FCW) modes, with an applied field $H$ = 50 Oe.

FIG. 3 Temperature dependence of ac susceptibility for LCMCO system

FIG. 4 Magnetization isotherms at selected temperatures, which are labeled with "^" in figure 3 at (a) 10 K, (b) 30 K, (c) 70 K, (d) 120 K, (e) 150 K, and (f) 200 K and 300 K, respectively.

FIG. 5 Magnetization isotherms at $T$ = 2.5 K.

Gao et al, FIG. 1

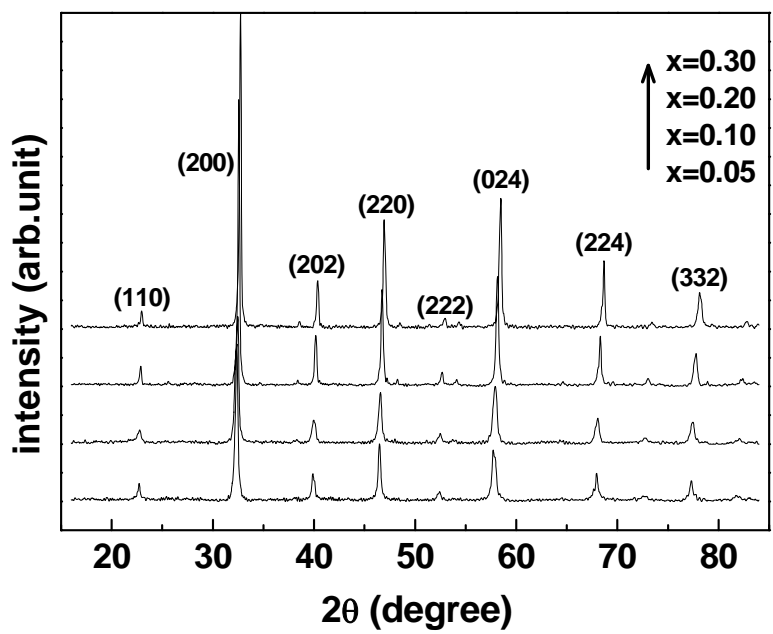

Gao et al, FIG. 2

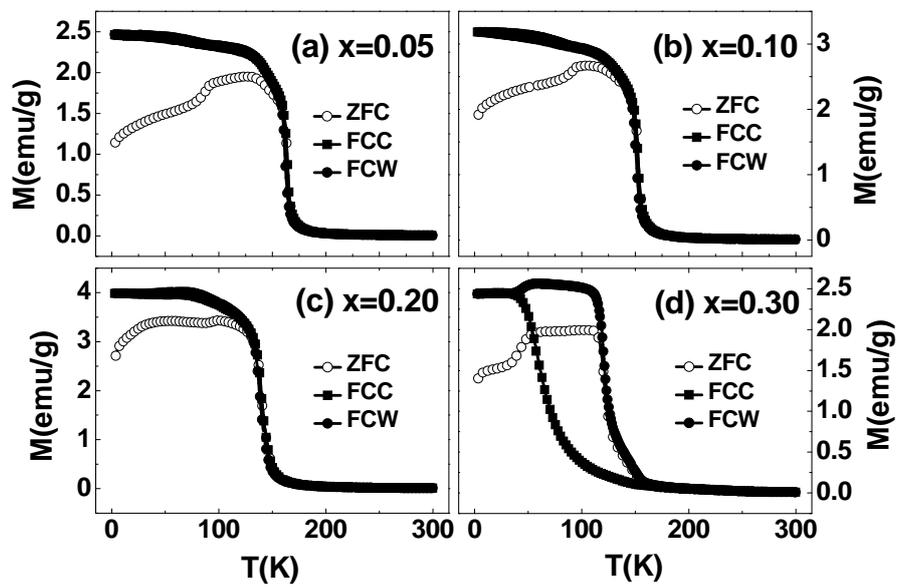

Gao et al, FIG. 3

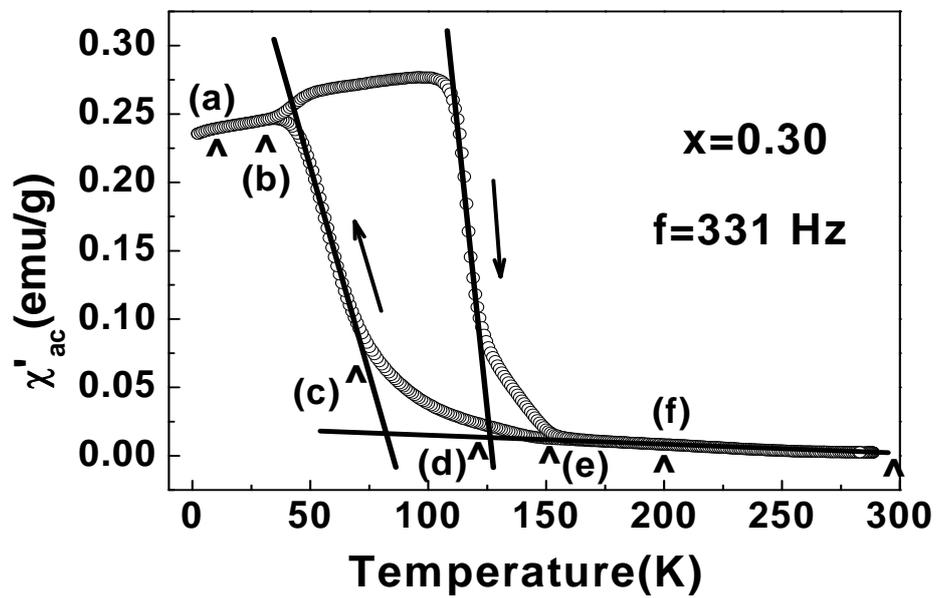

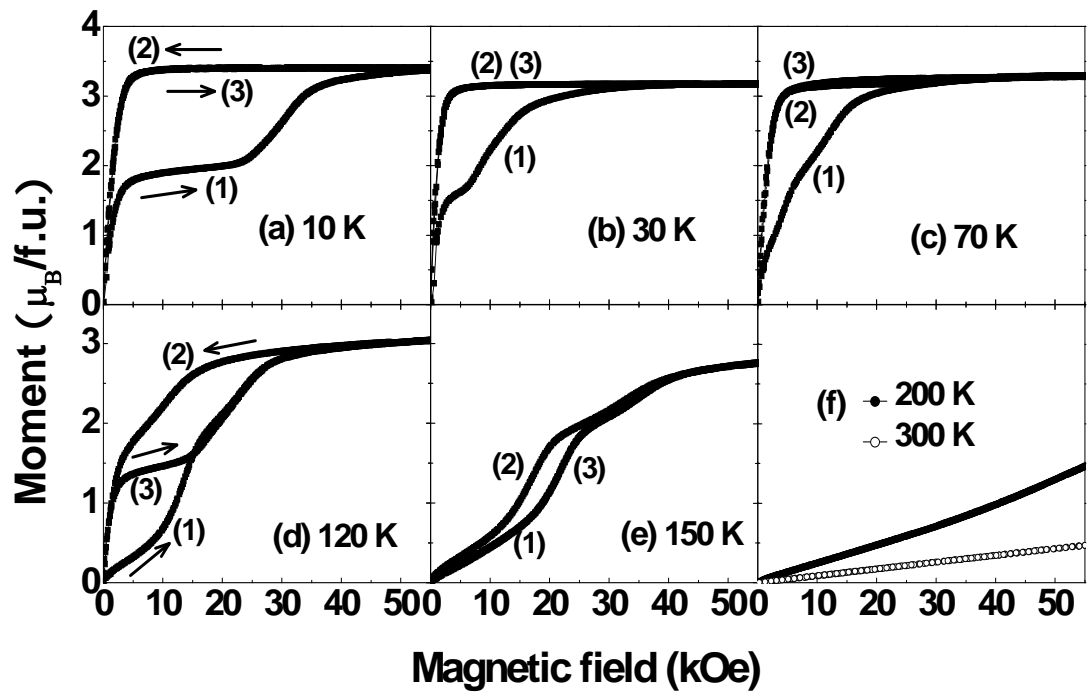

Gao et al, FIG. 5

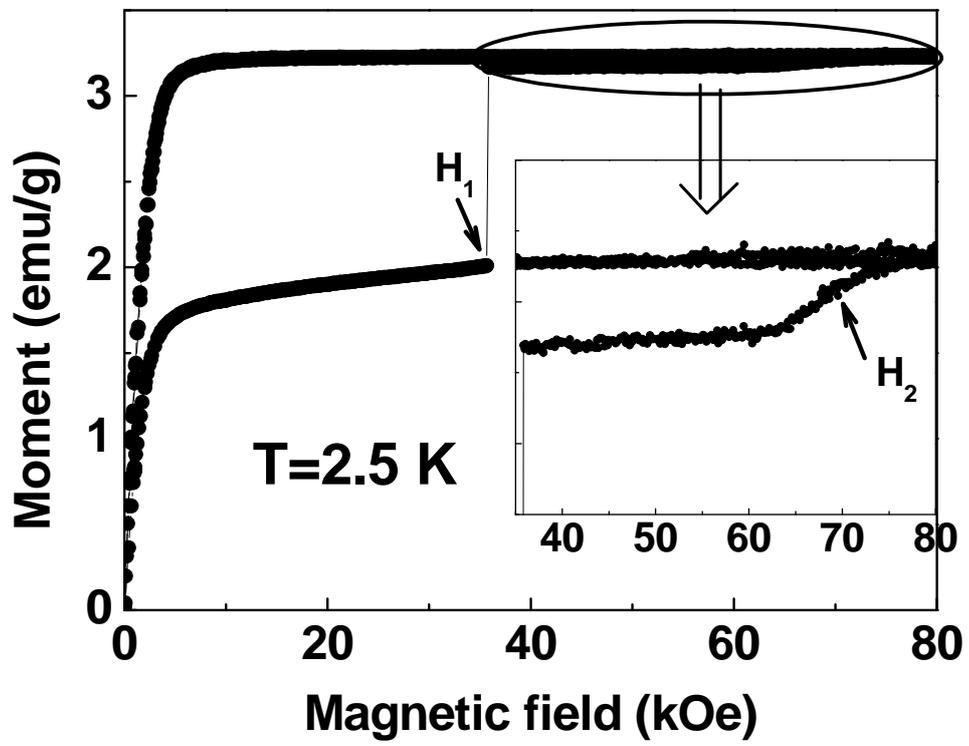